\documentclass[aps,prl,reprint,superscriptaddress,amsmath,amssymb,showpacs,floatfix]{revtex4-1}
\usepackage{graphicx}
\usepackage{bm}
\usepackage{color}
\usepackage{CJK}
\usepackage{xspace}
\newcommand{\be}{\begin{equation}}
\newcommand{\ee}{\end{equation}}
\newcommand{\xu}{\bar{u}}
\newcommand{\xd}{u}
\newcommand{\vu}{\bar{v}}
\newcommand{\vd}{v}
\newcommand{\nameone}{MCM\xspace}

\begin{document}
\begin{CJK*}{UTF8}{mj}

\title{Coarsening Dynamics of Nonequilibrium Chiral Ising Models}

\author{Mina Kim (김민아)}
\affiliation{Department of Physics, University of Seoul, Seoul 130-743,
Korea}
\author{Su-Chan Park (박수찬)}
\affiliation{Department of Physics, The Catholic University of Korea,
Bucheon, 420-743, Korea}
\author{Jae Dong Noh (노재동)}
\affiliation{Department of Physics, University of Seoul, Seoul 130-743,
Korea}
\affiliation{School of Physics, Korea Institute for Advanced Study,
Seoul 130-722, Korea}
\date{\today}

\begin{abstract}
We investigate a nonequilibrium coarsening dynamics of a one-dimensional 
Ising spin system with chirality. Only spins at domain
boundaries are updated so that the model undergoes a coarsening to 
either of equivalent absorbing states with all spins $+$ or $-$. Chirality
is imposed by assigning different transition rates to events at down~($+-$) 
kinks and up~($-+$) kinks. 
The coarsening is characterized by power-law scalings of the kink
density $\rho \sim t^{-\delta}$ and the characteristic length scale
$\xi \sim t^{1/z}$ with time $t$. Surprisingly the scaling exponents vary
continuously with model parameters, which is not the case for systems
without chirality. These results are obtained from extensive 
Monte Carlo simulations and spectral analyses of the time evolution
operator. 
Our study uncovers the novel universality 
class of the coarsening dynamics with chirality. 
\end{abstract}
\pacs{02.50.Ey, 05.50.+q, 05.70.Ln}

\maketitle
\end{CJK*}

Coarsening takes place in various systems such as magnetic systems, binary
alloys, and social systems with opinion dynamics. 
When a system is quenched from a high-temperature disordered phase
to a low-temperature ordered phase, a typical size of domains
grows in time following a power law
\be\label{z_def}
\xi \sim t^{1/z} \ ,
\ee
with dynamic exponent $z$.
It is known that coarsening systems 
are classified into a few universality classes depending on spatial 
dimensionality, order parameter symmetry, conservation in dynamics, 
and so on~\cite{Bray02}. 

The Ising model is one of the best studied coarsening systems.
It is symmetric under the global spin inversion~($Z_2$ symmetry) and has a
scalar order parameter. 
Under the single-spin-flip Glauber dynamics~\cite{Glauber63} that does not
conserve the order parameter, the coarsening dynamics is characterized
by $z=2$. On the other hand, the dynamic exponent is given by $z=3$
under the spin-exchange Kawasaki dynamics~\cite{Kawasaki66} conserving
the order parameter. Systems with nonscalar order parameter constitute
distinct universality classes~\cite{Bray02}.

A coarsening process is rather simple in systems with
discrete symmetry and nonconserving dynamics in one dimension. 
Consider a one-dimensional~(1D) Ising spin chain  with the Glauber dynamics 
at zero temperature, or equivalently the voter model~\cite{Liggett95}. 
In this model, only spins at domain boundaries can flip so
that domain walls diffuse and
annihilate in pairs. The diffusive nature suggests that the dynamic exponent
is given by $z=2$ and that the domain wall density decays algebraically 
as 
\be\label{delta_def}
\rho \sim t^{-\delta}
\ee
with an exponent $\delta = 1/2$. These scaling laws are verified by 
the exact solution~\cite{Glauber63,Cox89,Amar90}.

The power-law scaling with $z=2$ and $\delta=1/2$ seems to be robust in one
dimension. 
The $q$-state Potts model with the zero-temperature Glauber dynamics 
exhibits the same scaling behavior~\cite{Sire95,Derrida95,Derrida96,Cox89}.
It is also observed in nonequilibrium systems. 
Consider the voter model with an additional exchange process of 
neighboring spins~\cite{Dornic01}. 
It is equivalent to the branching annihilating random walk~(BAW) 
model~\cite{Menyhard94,TT1992}, where domain walls diffuse, annihilate 
in pairs, and branch two offsprings.
Despite the branching, the model displays the coarsening with the same 
exponents~\cite{TT1992,bALR1994}. The voter model with a kinetic constraint
also displays the same scaling behavior with a logarithmic
correction~\cite{MDG2001}.

Most studies on the coarsening have focused on
the role of order parameter symmetry~\cite{Bray02}. 
On the other hand, some dynamical systems are
characterized by coupled symmetry and little is known about the
coarsening dynamics in such systems.
In this Letter, we investigate the coarsening dynamics of 
a 1D Ising spin system which is invariant under the simultaneous inversion
of spin and space. Remarkably, the model with the coupled symmetry 
constitutes a novel universality class that is
characterized by continuously varying exponents.

Our study was initially motivated by a flocking phenomenon of self-propelled 
particles.
Flocking means a collective motion of particles, which is easily observed
in a group of birds, fish, insects, or animals in nature. Vicsek {\em et
al}~\cite{Vicsek95} proposed a simple model where the motion of each particle 
$i$ at position $\bm{r}_i$ is described by a velocity vector $\bm{v}_i$ of 
a constant speed.  Each time step, the direction 
of $\bm{v}_i$ is updated to the average direction of particles within a
fixed distance perturbed by a random noise. 
The system coarsens into a flocking phase when the noise strength is 
small~\cite{Vicsek95,Gregoire04}. 
Note that spatial isotropy is broken spontaneously due to the motion of particles. Consequently,
the model is symmetric under the simultaneous rotation/inversion of 
the velocity and the space.

In order to investigate the effect of coupled symmetry, 
we study an Ising spin system $\{\bm{\sigma} = (\sigma_1,\cdots,\sigma_N)\}$ 
with $\sigma_i=\pm 1$ in a 1D lattice of 
$N$ sites under periodic boundary conditions. 
Only spins near domain walls or kinks are updated following the rule
\begin{eqnarray}
+- \stackrel{\xd}{\longrightarrow} -+,\quad\quad 
&&-+ \stackrel{\xu}{\longrightarrow} +-,\nonumber\\
+- \stackrel{\vd/2}{\longrightarrow} \begin{cases} ++, \\ --, \end{cases}
\quad
&&-+ \stackrel{\vu/2}{\longrightarrow} \begin{cases} ++, \\ --, \end{cases}
\end{eqnarray}
where $+$ ($-$) stands for a spin with value $+1$ ($-1$) and parameters 
over the arrows denote the transition rates of 
corresponding events. Without losing generality, we set 
$\xd + \vd = 1$ and $\xu + \vu \le 1$.
The model may be represented with two kinds of domain walls or
kinks: $A$ and $B$ corresponding to a {\em down kink} $+-$ and an {\em up
kink} $-+$, respectively. The spin dynamics are then translated to an unbiased
hopping of $A$~($B$) with rate $\vd$~($\vu$) and branching of $A\to
ABA$~($B\to BAB$) with rate $\xd$~($\xu$). 
Opposite kinks annihilate in pair upon collision. 

\begin{figure}[t]
\includegraphics*[width=\columnwidth]{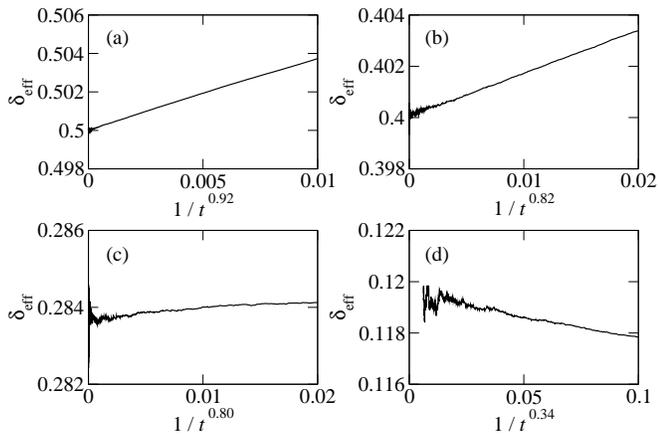}
\caption{\label{fig1} Plots of effective
exponents $\delta_\textrm{eff}$
for (a) $\xd =$ 0, (b) 0.3, (c) 0.6, and (d) 0.9. 
}
\end{figure}
Our model is characterized by {\em chirality}: The transition rates 
for events associated with down and up kinks are different. 
This chirality breaks the $Z_2$ symmetry, but leaves system invariant under
the simultaneous inversion of spin and space, $\sigma_i \to -\sigma_{-i}$. 
Emphasizing the role of the chirality, the model will be referred to as the 
nonequilibrium chiral Ising model~(NCIM). 
We remark that chirality is irrelevant for equilibrium Ising
systems~\cite{Ostlund81}. However, it turns out to result in an interesting 
feature in nonequilibrium cases.

The NCIM reduces to the voter model when $\vd=\vu$ and 
$\xd=\xu=0$, and the asymmetric simple exclusion
process~(ASEP)~\cite{Liggett95} when $\vd=\vu=0$.
A mixture of them was studied in Refs.~\cite{Belitsky01,MacPhee10} and 
was found to display complicated scaling behaviors.
When $\xd = \xu$ and $\vd = \vu$ ($Z_2$ symmetric case without chirality), 
discerning $A$ from $B$ becomes nominal and 
the model becomes equivalent to the BAW model~\cite{TT1992}.
It is solvable exactly~\cite{bALR1994} 
and the kink density decays with $\delta =1/2$ for all $\xd<1$.
Note that the NCIM is different from the so-called directed Ising model 
in which kinks are biased to a preferred 
direction~\cite{Godreche09,Godreche11}.

First, we study the maximum chirality case with $\xu = \vu = 0$ where up
kinks~($B$) are immobile and do not branch offsprings. 
This case is analogous to a 1D version of the flocking model~\cite{Vicsek95}
if $+$ ($-$) spin is interpreted as a bird moving to the right (left) and if
a bird is assumed to interact with another bird along 
its moving direction. 
For convenience, we refer to this case as the maximum chirality 
model~(\nameone).

The system is prepared in an anti-ferromagnetic state $(\cdots +-+-\cdots)$ 
initially. Then, we measure the total kink density and average over 
$N_S$ samples to obtain $\rho(t)$ for $t\leq 10^7$.
Just like the BAW model, we expect that 
$\rho(t)$ decays in a power-law fashion for all $\xd<1$~\cite{supplemental}. 
So we investigate
the behavior of an effective exponent defined as 
\be
\label{delta_eff}
\delta_\text{eff}(t) = -\log\left[\rho(t)/\rho(t/b)\right]/\log{b}
\ee
with a constant $b$.
If the long time behavior of the kink density
is given by $\rho(t) = t^{-\delta}( a + c t^{-\zeta})$ 
up to a leading correction to scaling, the effective exponent in the long
time limit behaves as
$ \delta_\text{eff}(t) \simeq \delta + a_1 t^{-\zeta}, $
where $a_1$ is a constant depending on $b$.
So when we draw $\delta_\text{eff}$ against $t^{-\zeta}$ 
with the correct value of $\zeta$, $\delta_\text{eff}(t)$ should approach to 
$\delta$ with a finite slope as $t\rightarrow \infty$.

\begin{table}[t]
\caption{\label{table1} Numerical values of $\delta$ and $z$ of the \nameone for
various values of $u$. For $z$, we present
the results from the eigenspectrum analysis [$z$~(spectrum)] and
from the seed simulations [$z$~(seed)]. The numbers
in parentheses indicate errors of the last digits. }
\begin{ruledtabular}
\begin{tabular}{llll}
$u$ & $\delta$ & $z$~(spectrum) & $z$~(seed)\\
\hline
0   & 0.5000(1) &2.0000(6)  &1.998(8)    \\
0.1 & 0.4678(2) &1.8789(5)  &1.877(9) \\
0.2 & 0.4346(3) &1.7683(6)  &1.769(4) \\
0.3 & 0.4001(2) &1.6666(6)  &1.667(2) \\
0.4 & 0.3639(5) &1.5716(8)  &1.570(6) \\
0.5 & 0.3254(5) &1.4818(7)  &1.483(4) \\
0.6 & 0.2837(4) &1.3958(4)  &1.397(4) \\
0.7 & 0.2376(8) &1.3117(8)  &1.313(6) \\
0.8 & 0.1850(8) &1.227(2)   &1.228(3) \\
0.9 & 0.1195(3) &1.136(5)   &1.139(6)
\end{tabular}
\end{ruledtabular}
\end{table}
Figure~\ref{fig1} presents the behavior of the effective exponents
for $u=0$, 0.3, 0.6, and 0.9. The number of runs for all $u$ is $N_S = 5000$
and system sizes for $u=0$, 0.3, 0.6, and 0.9 are $N=2^{24}$, $2^{23}$,
$2^{22}$, and $2^{21}$ , respectively.
These system sizes 
are large enough that all runs did not fall into absorbing states
where all spins have the same sign. 

\begin{figure}[t]
\includegraphics*[width=\columnwidth]{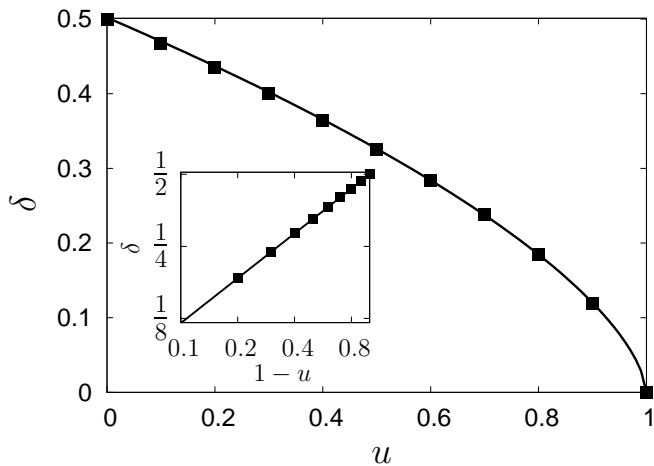}
\caption{\label{fig2} Plots of $\delta$ vs $u$ (symbols) and its fitting 
function $a_2(1-u)^\chi$ with $a_2\approx 0.5$ and 
$\chi \approx 0.62$ (lines). 
Inset: Log-log plot of $\delta$ against $1-u$ together with the fitting 
function.}
\end{figure}

The correction-to-scaling exponent $\zeta$ are roughly estimated
from a fitting of $\delta_\text{eff}$ to the from $\delta + a_1 t^{-\zeta}$
with three fitting parameters $\delta$, $a_1$, and $\zeta$.
This procedure can be error-prone and the accuracy of the estimated
$\zeta$ may be questionable. Still, however, 
if we draw $\delta_\text{eff}$ against $t^{-\zeta}$ with $\zeta$ from a fitting, 
we observed that $\delta_\text{eff}$'s intersect the $y$-axis with 
finite slope, which allows for estimating $\delta$.
We observed that the corrections to scaling become stronger 
as $\xd$ approaches 1, which is thought to be
a signature of the crossover from the mean-field 
voter dynamics to the \nameone; see below.
The analysis of the effective exponent for other values of $u$ is presented in
Supplemental Material~\cite{supplemental}.
Table~\ref{table1} summarizes numerical results of 
$\delta$ for various $u$'s and
Fig.~\ref{fig2} illustrates $\delta$ against $u$. 

The intriguing feature of the \nameone is that the exponent $\delta$ varies 
with $\xd$. Furthermore, as $u$ approaches 1,
$\delta$ seems to show a non-trivial power-law behavior. Indeed, if we fit 
$\delta$ for the region $u \ge 0.6$ using 
$ \delta \approx a_2 (1-u)^\chi$,  
with two fitting parameters $a_2$ and $\chi$, we found that it fits the data
quite well with $a_2 \approx 0.5$ and $\chi \approx 0.62$; see Fig.~\ref{fig2}. 
This singular behavior of 
$\delta$ at $u=1$ can be understood as a result of
a crossover from the mean-field voter dynamics to the \nameone.

When $u$ is very close to 1 but not exactly 1, the voter dynamics which 
occurs with rate $v=1-u$ happens after many attempts of the ASEP dynamics.
Since the stationary state of the ASEP is totally
uncorrelated~\cite{Liggett95}, the voter dynamics can happen only after 
all spins are distributed almost randomly. 
Hence, if we rescale the time as $\tau =  (1-u)t$ and take
a limit $u\rightarrow 1$ with $\tau$ kept finite, the \nameone should be the 
same as a mean-field voter model on a complete graph. 
Since coarsening does not occur
on a complete graph of infinite size, $\delta$ should be zero
in the above mentioned limit. Thus, there should be a crossover from the
mean-field voter dynamics to the 1D \nameone at $u=1$.
Note that the above argument does not depend
on whether $\vd$ and/or $\vu$ are zero or not once $\vu$ approaches 
zero as $u\rightarrow 1$. 
Although the accuracy of $\chi$ may be questionable, we can still insist that 
the crossover is described by the (non-trivial) exponent $\chi$.

We substantiate the Monte Carlo results by studying the spectrum 
of the time evolution operator of the MCM. 
In general, a master equation can be mapped to 
an imaginary time Schr\"odinger equation
with {\it Hamiltonian} $\mathsf{H}$ whose 
eigenvalues contain most of relevant information of the system. 
For instance, the directed percolation system has been studied
successfully with the eigenspectrum 
analysis~\cite{Henkel90,BenAvraham91,Carlon99}.

For the \nameone, the Hamiltonian takes the form
\begin{eqnarray}
\mathsf{H} &=&  
   \frac{1}{4}\sum_{i=1}^N (1+\hat\sigma_i^z)(1-\hat\sigma_{i+1}^z) 
   -\xd\sum_{i=1}^N \hat\sigma_i^-\hat\sigma_{i+1}^+ \nonumber \\
   && - \frac{(1-\xd)}{4} \sum_{i=1}^N
\left\{ \hat\sigma_i^- (1 - \hat\sigma_{i+1}^z) +
(1+\hat\sigma_i^z) \hat\sigma_{i+1}^+\right\} ,
\end{eqnarray}
where $\hat\sigma_i$ is the Pauli spin operator acting on a spin at site $i$. 
We label eigenvalues of $\mathsf{H}$ as $E_n$  with $n=1,\cdots,2^N$ and 
call $E_n$ the energy of the $n$-th
level. Since $\mathsf{H}$ is not Hermitian, $E_n$ may have a complex value.
The eigenvalues are sorted in the ascending order of $\Re[E_n]$, 
the real part of $E_n$.
There are two trivial levels with $E_1=E_2=0$ 
corresponding to the two absorbing states with all spins having the same
sign.  Other low-lying energy levels with $n>2$ define the relaxation time 
as $ \tau_n = {1}/({\Re[E_n]})$. 

We have diagonalized numerically the Hamiltonian up to $N=20$ to
obtain the longest relaxation time $\tau_3$.
Since $\tau_3 \sim N^z$, the dynamic exponent $z$ is estimated by
extrapolating an effective exponent $z_\text{eff}(N) \equiv 
\ln[\tau_3 (N)/\tau_3 (N-2)] / \ln [N/(N-2)]$ with the Bulirsch-Stoer~(BST)
algorithm~\cite{BS1964,Henkel88}. 
The results are summarized in Table~\ref{table1}.
As an illustration, detailed analysis of $z_\text{eff}$ for $u=0.8$
along with the extrapolation according to the BST algorithm
is presented in Supplemental Material~\cite{supplemental}.

We have also performed independent Monte Carlo simulations starting 
with a single down kink $A$ in the middle of an infinite lattice, which
are generally referred to as seed simulations. We have measured the
particle spreading distance $\xi(t)\sim t^{1/z}$ to obtain the dynamic
exponent $z$ following the similar effective exponent analysis as done for
$\delta$. Both results for $z$ are in perfect agreement with each other
and vary continuously with $\xd$. 
See Supplemental Material~\cite{supplemental} for the full analyses of $z$.

We have shown that the coarsening dynamics of the \nameone is characterized by 
continuously varying critical exponents. 
Note that chirality lies both in the ASEP events~($\xd\neq
\xu$) and the voter-model events~($\vd\neq\vu$).
In order to investigate which one is the essential ingredient, 
we studied two more cases: One is the case with
$\xd = \xu$ and $\vd \neq \vu = 0$ which will be called the
symmetric exclusion and chiral voter (SECV) model and the other is
the case with $\xd\neq \xu=0$ and $\vd = \vu$ to be called
the chiral exclusion and symmetric voter (CESV) model. The CESV model is a
particular limiting case of the model studied in
Ref.~\cite{Belitsky01,MacPhee10}.
Remind that $\vd$ is always set to $1-\xd$.

\begin{figure}[t]
\includegraphics*[width=\columnwidth]{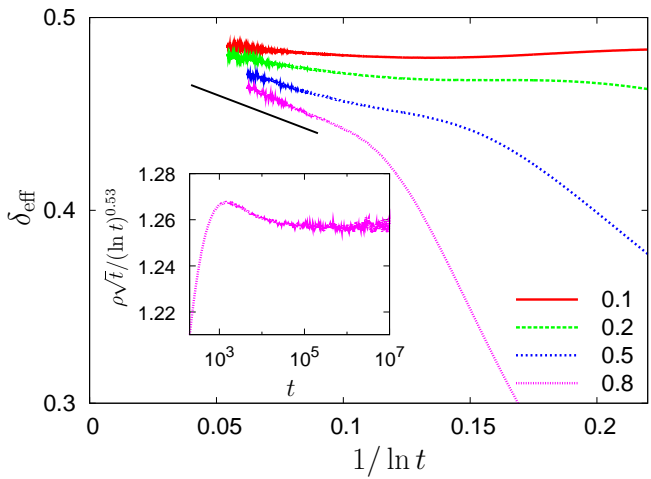}
\caption{\label{fig:SECV} (Color online) Plots of $\delta_\text{eff}$ as functions
of $1/\ln t $ for $\xd = 0.1$, 0.2, 0.5, and 0.8 (from top to bottom) 
for the SECV. A line segment with slope $-0.5$ is for a guide
to the eyes. Inset: Plot of $\rho(t) \sqrt{t}/(\ln t)^{0.53}$ vs $t$ for the SECV
with $\xd = 0.8$ on a semi-logarithmic scale.}
\end{figure}
The analyses of $\delta_\text{eff}$ for the SECV with 
$\xd=0.1$, 0.2, 0.5, and 0.8 are summarized in 
Fig.~\ref{fig:SECV}. It seems that $\delta_\text{eff}(t)$ for all
$\xd$ approaches $1/2$ with logarithmic corrections.  
Notice that if there exists a logarithmic correction as
\be
\rho(t) \sim (\ln t + C)^\kappa / t^\delta
\label{log_correction}
\ee 
with a constant $C$,
the effective exponent defined in Eq.~(\ref{delta_eff}) should behave as
$\delta_\text{eff}(t)  \approx \delta - {\kappa}/{\ln t}$
in the asymptotic regime. 
Thus, if we plot $\delta_\text{eff}(t)$ as a function of 
$1/\ln t$, the effective exponent should intersect the $y$-axis with 
slope $-\kappa$.
This phenomenon is quite pronounced for the cases of $\xd = 0.5$ and $0.8$ 
and the slope seems to be around 0.5. 
Indeed, if $\rho(t) \sqrt{t}/(\ln t)^{0.53}$ 
is plotted against $t$ on a semi-logarithmic scale 
(see Inset of Fig.~\ref{fig:SECV} for the case of $\xd =0.8$), 
a flat region is observable in the long
time limit for more than two log-decades. 
Although the accurate value of $\kappa$ is hard to estimate, 
we can conclude that there exist a systematic logarithmic correction as
shown in Eq.~(\ref{log_correction}) with the leading scaling exponent
$\delta=1/2$ unchanged in the SECV model. 

A logarithmic correction in a 1D
coarsening has been reported in a different model~\cite{MDG2001} that
corresponds to the voter model with a weak kinetic constraint.
In that model, the kink density decays faster than
$1/\sqrt{t}$ as $1/(\sqrt{t} \ln t)$, but in our case
it decays slower than $1/\sqrt{t}$. 
Qualitatively, this slowing down should be attributed to the 
presence of branching dynamics which increases the number of kinks.
However, a quantitative analysis requires further investigation,
which is beyond the scope of this Letter.
For our purpose, it is enough to conclude that the continuously 
varying exponents are not due to the chiral voter dynamics.

\begin{figure}[t]
\includegraphics*[width=\columnwidth]{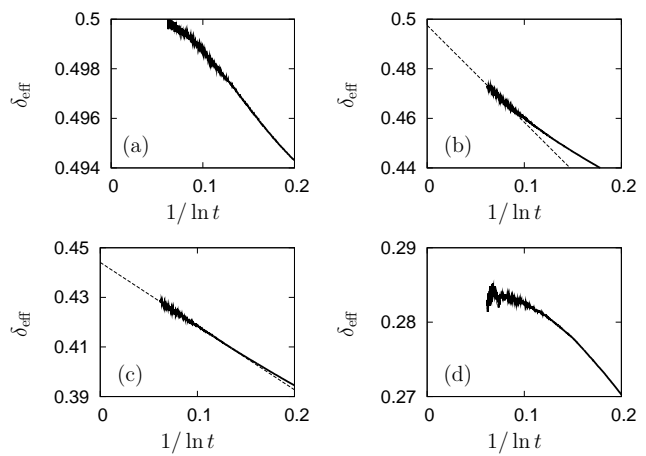}
\caption{\label{fig:CESV} Plots of $\delta_\text{eff}$ as functions
of $1/\ln(t)$ for $\xd$ = (a) 0.2, (b) 0.5, (c) 0.6, and (d) 0.8  
for the CESV. For the case of $\xd =$ 0.5 and 0.6 [(b) and (c)], fitting
results of a function $\delta + d/\ln t$ with $\delta$ and $d$ fitting
parameters are also depicted.}
\end{figure}

The CESV model shows a more intriguing feature. We present the effective
exponent data for the density decay in Fig.~\ref{fig:CESV}.
For $\xd = 0.2$~[Fig.~\ref{fig:CESV}(a)], 
$\delta_\text{eff}$ seems to approach 0.5 with negligible
logarithmic correction. For $\xd= 0.5$~[Fig.~\ref{fig:CESV}(b)], 
we cannot make a firm conclusion whether $\delta < 0.5$ or 
$\delta = 0.5$ due to strong correction-to-scaling behavior. 
Quite interestingly, when $\xd > 0.5$~[Fig.~\ref{fig:CESV}(c) and (d)], 
$\delta$ deviates from 0.5 significantly even under the assumption
of a logarithmic correction. So we conclude that the CESV has 
continuously varying exponents with possible logarithmic corrections
when $u > 0.5$. 
This study shows that the chirality in the spin exchange is responsible for
the continuously varying exponents. Nevertheless, it remains open
why and when there appears a logarithmic correction in the coarsening
process.

To summarize, we have studied the one-dimensional coarsening dynamics of
nonequilibrium Ising spin systems with chirality. Although chirality is
irrelevant in equilibrium Ising systems, it turns out that the chirality
can lead to continuously varying scaling exponents in the
nonequilibrium chiral Ising model.  In particular, it turns out that 
the chirality in spin exchange plays a crucial role. 

It is rare to observe continuously varying exponents
from systems without quenched disorder. 
The $q$-state Potts model with zero-temperature Glauber dynamics was 
studied in Refs.~\cite{Derrida95,Derrida96}. It was found that the critical
exponent describing the power-law decay of the persistent probability
varies continuously with $q$. Nevertheless, the coarsening dynamics 
is still pure diffusive and characterized by $z=2$ and $\delta=1/2$ at all
values of $q$.  
We notice that continuously varying exponents
were reported in a 1D sandpile model without
dissipation~\cite{Jain2005} and that there is actually a parallelism
between this sandpile model and the MCM. This connection will be 
discussed elsewhere~\cite{unpub}.

Some of Ising spin systems are exactly solvable in one 
dimension~\cite{Glauber63,Cox89,BenAvraham91,Derrida95}, for equations 
governing the time evolution of correlation functions are closed. In the 
presence of the chirality, however, the equations are not closed, which
makes the exact solution for the NCIM not available in general.
Our numerical finding of the universality class 
with continuously varying exponents can be established more firmly if one find 
a minimal continuum equation obeying the proper symmetry property. We leave it
as a future work~\cite{unpub}.

This work was supported by the National Research Foundation of Korea~(NRF)
grant funded by the Korea government~(MEST)~(No. 2012-0005003). S-CP
acknowledges the support by the Basic Science Research Program through the
National Research Foundation of Korea~(NRF) funded by the Ministry of
Education, Science and Technology~(Grant No. 2011-0014680). MK acknowledges
the financial support from the TJ Park Foundation. Discussion with J. Krug,
B. Derrida, and G. Sch\"utz is appreciated. The computation was partly
supported by Universit\"at zu K\"oln.

\bibliographystyle{apsrev}
\bibliography{ncim.bbl}

\begin{thebibliography}{29}
\expandafter\ifx\csname natexlab\endcsname\relax\def\natexlab#1{#1}\fi
\expandafter\ifx\csname bibnamefont\endcsname\relax
  \def\bibnamefont#1{#1}\fi
\expandafter\ifx\csname bibfnamefont\endcsname\relax
  \def\bibfnamefont#1{#1}\fi
\expandafter\ifx\csname citenamefont\endcsname\relax
  \def\citenamefont#1{#1}\fi
\expandafter\ifx\csname url\endcsname\relax
  \def\url#1{\texttt{#1}}\fi
\expandafter\ifx\csname urlprefix\endcsname\relax\def\urlprefix{URL }\fi
\providecommand{\bibinfo}[2]{#2}
\providecommand{\eprint}[2][]{\url{#2}}

\bibitem[{\citenamefont{Bray}(2002)}]{Bray02}
\bibinfo{author}{\bibfnamefont{A.~J.} \bibnamefont{Bray}},
  \bibinfo{journal}{Adv. in Phys.} \textbf{\bibinfo{volume}{51}},
  \bibinfo{pages}{481} (\bibinfo{year}{2002}).

\bibitem[{\citenamefont{Glauber}(1963)}]{Glauber63}
\bibinfo{author}{\bibfnamefont{R.~J.} \bibnamefont{Glauber}},
  \bibinfo{journal}{J. Math. Phys.} \textbf{\bibinfo{volume}{4}},
  \bibinfo{pages}{294} (\bibinfo{year}{1963}).

\bibitem[{\citenamefont{Kawasaki}(1966)}]{Kawasaki66}
\bibinfo{author}{\bibfnamefont{K.}~\bibnamefont{Kawasaki}},
  \bibinfo{journal}{Phys. Rev.} \textbf{\bibinfo{volume}{145}},
  \bibinfo{pages}{224} (\bibinfo{year}{1966}).

\bibitem[{\citenamefont{Liggett}(1995)}]{Liggett95}
\bibinfo{author}{\bibfnamefont{T.~M.} \bibnamefont{Liggett}},
  \emph{\bibinfo{title}{Interacting Particle Systems}}
  (\bibinfo{publisher}{Springer-Verlag}, \bibinfo{address}{New York},
  \bibinfo{year}{1995}).

\bibitem[{\citenamefont{Cox}(1989)}]{Cox89}
\bibinfo{author}{\bibfnamefont{J.~T.} \bibnamefont{Cox}},
  \bibinfo{journal}{Ann. of Probab.} \textbf{\bibinfo{volume}{17}},
  \bibinfo{pages}{1333} (\bibinfo{year}{1989}).

\bibitem[{\citenamefont{Amar and Family}(1990)}]{Amar90}
\bibinfo{author}{\bibfnamefont{J.~G.} \bibnamefont{Amar}} \bibnamefont{and}
  \bibinfo{author}{\bibfnamefont{F.}~\bibnamefont{Family}},
  \bibinfo{journal}{Phys. Rev. A} \textbf{\bibinfo{volume}{41}},
  \bibinfo{pages}{3258} (\bibinfo{year}{1990}).

\bibitem[{\citenamefont{Sire and Majumdar}(1995)}]{Sire95}
\bibinfo{author}{\bibfnamefont{C.}~\bibnamefont{Sire}} \bibnamefont{and}
  \bibinfo{author}{\bibfnamefont{S.~N.} \bibnamefont{Majumdar}},
  \bibinfo{journal}{Phys. Rev. E} \textbf{\bibinfo{volume}{52}},
  \bibinfo{pages}{244} (\bibinfo{year}{1995}).

\bibitem[{\citenamefont{Derrida et~al.}(1995)\citenamefont{Derrida, Hakim, and
  Pasquier}}]{Derrida95}
\bibinfo{author}{\bibfnamefont{B.}~\bibnamefont{Derrida}},
  \bibinfo{author}{\bibfnamefont{V.}~\bibnamefont{Hakim}}, \bibnamefont{and}
  \bibinfo{author}{\bibfnamefont{V.}~\bibnamefont{Pasquier}},
  \bibinfo{journal}{Phys. Rev. Lett.} \textbf{\bibinfo{volume}{75}},
  \bibinfo{pages}{751} (\bibinfo{year}{1995}).

\bibitem[{\citenamefont{Derrida and Zeitak}(1996)}]{Derrida96}
\bibinfo{author}{\bibfnamefont{B.}~\bibnamefont{Derrida}} \bibnamefont{and}
  \bibinfo{author}{\bibfnamefont{R.}~\bibnamefont{Zeitak}},
  \bibinfo{journal}{Phys. Rev. E} \textbf{\bibinfo{volume}{54}},
  \bibinfo{pages}{2513} (\bibinfo{year}{1996}).

\bibitem[{\citenamefont{Dornic et~al.}(2001)\citenamefont{Dornic, Chat{\'e},
  Chave, and Hinrichsen}}]{Dornic01}
\bibinfo{author}{\bibfnamefont{I.}~\bibnamefont{Dornic}},
  \bibinfo{author}{\bibfnamefont{H.}~\bibnamefont{Chat{\'e}}},
  \bibinfo{author}{\bibfnamefont{J.}~\bibnamefont{Chave}}, \bibnamefont{and}
  \bibinfo{author}{\bibfnamefont{H.}~\bibnamefont{Hinrichsen}},
  \bibinfo{journal}{Phys. Rev. Lett.} \textbf{\bibinfo{volume}{87}},
  \bibinfo{pages}{045701} (\bibinfo{year}{2001}).

\bibitem[{\citenamefont{Menyh\'ard}(1995)}]{Menyhard94}
\bibinfo{author}{\bibfnamefont{N.}~\bibnamefont{Menyh\'ard}},
  \bibinfo{journal}{J. Phys. A} \textbf{\bibinfo{volume}{27}},
  \bibinfo{pages}{6139} (\bibinfo{year}{1995}).

\bibitem[{\citenamefont{Takayasu and Tretyakov}(1992)}]{TT1992}
\bibinfo{author}{\bibfnamefont{H.}~\bibnamefont{Takayasu}} \bibnamefont{and}
  \bibinfo{author}{\bibfnamefont{A.~Y.} \bibnamefont{Tretyakov}},
  \bibinfo{journal}{Phys. Rev. Lett.} \textbf{\bibinfo{volume}{68}},
  \bibinfo{pages}{3060} (\bibinfo{year}{1992}).

\bibitem[{\citenamefont{ben Avraham et~al.}(1994)\citenamefont{ben Avraham,
  Leyvraz, and Redner}}]{bALR1994}
\bibinfo{author}{\bibfnamefont{D.}~\bibnamefont{ben-Avraham}},
  \bibinfo{author}{\bibfnamefont{F.}~\bibnamefont{Leyvraz}}, \bibnamefont{and}
  \bibinfo{author}{\bibfnamefont{S.}~\bibnamefont{Redner}},
  \bibinfo{journal}{Phys. Rev. E} \textbf{\bibinfo{volume}{50}},
  \bibinfo{pages}{1843} (\bibinfo{year}{1994}).

\bibitem[{\citenamefont{Majumdar et~al.}(2001)\citenamefont{Majumdar, Dean, and
  Grassberger}}]{MDG2001}
\bibinfo{author}{\bibfnamefont{S.~N.} \bibnamefont{Majumdar}},
  \bibinfo{author}{\bibfnamefont{D.~S.} \bibnamefont{Dean}}, \bibnamefont{and}
  \bibinfo{author}{\bibfnamefont{P.}~\bibnamefont{Grassberger}},
  \bibinfo{journal}{Phys. Rev. Lett.} \textbf{\bibinfo{volume}{86}},
  \bibinfo{pages}{2301} (\bibinfo{year}{2001}).

\bibitem[{\citenamefont{Vicsek et~al.}(1995)\citenamefont{Vicsek, Czir{\'ok},
  Ben-Jacob, Cohen, and Shochet}}]{Vicsek95}
\bibinfo{author}{\bibfnamefont{T.}~\bibnamefont{Vicsek}},
  \bibinfo{author}{\bibfnamefont{A.}~\bibnamefont{Czir{\'ok}}},
  \bibinfo{author}{\bibfnamefont{E.}~\bibnamefont{Ben-Jacob}},
  \bibinfo{author}{\bibfnamefont{I.}~\bibnamefont{Cohen}}, \bibnamefont{and}
  \bibinfo{author}{\bibfnamefont{O.}~\bibnamefont{Shochet}},
  \bibinfo{journal}{Phys. Rev. Lett.} \textbf{\bibinfo{volume}{75}},
  \bibinfo{pages}{1226} (\bibinfo{year}{1995}).

\bibitem[{\citenamefont{Gr{\'e}goire and Chat{\'e}}(2004)}]{Gregoire04}
\bibinfo{author}{\bibfnamefont{G.}~\bibnamefont{Gr{\'e}goire}}
  \bibnamefont{and}
  \bibinfo{author}{\bibfnamefont{H.}~\bibnamefont{Chat{\'e}}},
  \bibinfo{journal}{Phys. Rev. Lett.} \textbf{\bibinfo{volume}{92}},
  \bibinfo{pages}{025702} (\bibinfo{year}{2004}).

\bibitem[{\citenamefont{Ostlund}(1981)}]{Ostlund81}
\bibinfo{author}{\bibfnamefont{S.}~\bibnamefont{Ostlund}},
  \bibinfo{journal}{Phys. Rev. B} \textbf{\bibinfo{volume}{24}},
  \bibinfo{pages}{398} (\bibinfo{year}{1981}).

\bibitem[{\citenamefont{Belitsky et~al.}(2001)\citenamefont{Belitsky, Ferrari,
  Menshikov, and Popov}}]{Belitsky01}
\bibinfo{author}{\bibfnamefont{V.}~\bibnamefont{Belitsky}},
  \bibinfo{author}{\bibfnamefont{P.~A.} \bibnamefont{Ferrari}},
  \bibinfo{author}{\bibfnamefont{M.~V.} \bibnamefont{Menshikov}},
  \bibnamefont{and} \bibinfo{author}{\bibfnamefont{S.~Y.} \bibnamefont{Popov}},
  \bibinfo{journal}{Bernoulli} \textbf{\bibinfo{volume}{7}},
  \bibinfo{pages}{119} (\bibinfo{year}{2001}).

\bibitem[{\citenamefont{MacPhee et~al.}(2010)\citenamefont{MacPhee, Menshikov,
  Volkov, and Wade}}]{MacPhee10}
\bibinfo{author}{\bibfnamefont{I.~M.} \bibnamefont{MacPhee}},
  \bibinfo{author}{\bibfnamefont{M.~V.} \bibnamefont{Menshikov}},
  \bibinfo{author}{\bibfnamefont{S.}~\bibnamefont{Volkov}}, \bibnamefont{and}
  \bibinfo{author}{\bibfnamefont{A.~R.} \bibnamefont{Wade}},
  \bibinfo{journal}{Bernoulli} \textbf{\bibinfo{volume}{16}},
  \bibinfo{pages}{1312} (\bibinfo{year}{2010}).

\bibitem[{\citenamefont{Godr{\`e}che and Bray}(2009)}]{Godreche09}
\bibinfo{author}{\bibfnamefont{C.}~\bibnamefont{Godr{\`e}che}}
  \bibnamefont{and} \bibinfo{author}{\bibfnamefont{A.~J.} \bibnamefont{Bray}},
  \bibinfo{journal}{J. Stat. Mech.} \textbf{\bibinfo{volume}{2009}},
  \bibinfo{pages}{P12016} (\bibinfo{year}{2009}).

\bibitem[{\citenamefont{Godr{\`e}che}(2011)}]{Godreche11}
\bibinfo{author}{\bibfnamefont{C.}~\bibnamefont{Godr{\`e}che}},
  \bibinfo{journal}{J. Stat. Mech.} \textbf{\bibinfo{volume}{2011}},
  \bibinfo{pages}{P04005} (\bibinfo{year}{2011}).

\bibitem[{sup()}]{supplemental}
\bibinfo{journal}{See Supplemental Material at URL for more details.} 

\bibitem[{\citenamefont{Henkel and Herrmann}(1990)}]{Henkel90}
\bibinfo{author}{\bibfnamefont{M.}~\bibnamefont{Henkel}} \bibnamefont{and}
  \bibinfo{author}{\bibfnamefont{H.~J.} \bibnamefont{Herrmann}},
  \bibinfo{journal}{J. Phys. A} \textbf{\bibinfo{volume}{23}},
  \bibinfo{pages}{3719} (\bibinfo{year}{1990}).

\bibitem[{\citenamefont{ben Avraham et~al.}(1991)\citenamefont{ben Avraham,
  Bidaux, and Schulman}}]{BenAvraham91}
\bibinfo{author}{\bibfnamefont{D.}~\bibnamefont{ben-Avraham}},
  \bibinfo{author}{\bibfnamefont{R.}~\bibnamefont{Bidaux}}, \bibnamefont{and}
  \bibinfo{author}{\bibfnamefont{L.~S.} \bibnamefont{Schulman}},
  \bibinfo{journal}{Phys. Rev. A} \textbf{\bibinfo{volume}{43}},
  \bibinfo{pages}{7093} (\bibinfo{year}{1991}).

\bibitem[{\citenamefont{Carlon et~al.}(1999)\citenamefont{Carlon, Henkel, and
  Schollw{\"o}ck}}]{Carlon99}
\bibinfo{author}{\bibfnamefont{E.}~\bibnamefont{Carlon}},
  \bibinfo{author}{\bibfnamefont{M.}~\bibnamefont{Henkel}}, \bibnamefont{and}
  \bibinfo{author}{\bibfnamefont{U.}~\bibnamefont{Schollw{\"o}ck}},
  \bibinfo{journal}{Eur. Phys. J. B} \textbf{\bibinfo{volume}{12}},
  \bibinfo{pages}{99} (\bibinfo{year}{1999}).

\bibitem[{\citenamefont{Bulirsch and Stoer}(1964)}]{BS1964}
\bibinfo{author}{\bibfnamefont{R.}~\bibnamefont{Bulirsch}} \bibnamefont{and}
  \bibinfo{author}{\bibfnamefont{J.}~\bibnamefont{Stoer}},
  \bibinfo{journal}{Numer. Math.} \textbf{\bibinfo{volume}{6}},
  \bibinfo{pages}{413} (\bibinfo{year}{1964}).

\bibitem[{\citenamefont{Henkel and Sch\"utz}(1988)}]{Henkel88}
\bibinfo{author}{\bibfnamefont{M.}~\bibnamefont{Henkel}} \bibnamefont{and}
  \bibinfo{author}{\bibfnamefont{G.}~\bibnamefont{Sch\"utz}},
  \bibinfo{journal}{J. Phys. A} \textbf{\bibinfo{volume}{21}},
  \bibinfo{pages}{2617} (\bibinfo{year}{1988}).

\bibitem[{\citenamefont{Jain}(2005)}]{Jain2005}
\bibinfo{author}{\bibfnamefont{K.}~\bibnamefont{Jain}},
  \bibinfo{journal}{Europhys. Lett.} \textbf{\bibinfo{volume}{71}},
  \bibinfo{pages}{8} (\bibinfo{year}{2005}).

\bibitem[{\citenamefont{Kim et~al.}(2012)\citenamefont{Kim, Noh, and
  Park}}]{unpub}
\bibinfo{author}{\bibfnamefont{M.}~\bibnamefont{Kim}},
  \bibinfo{author}{\bibfnamefont{J.~D.} \bibnamefont{Noh}}, \bibnamefont{and}
  \bibinfo{author}{\bibfnamefont{S.-C.} \bibnamefont{Park}}
  \bibinfo{journal}{(unpublished)}.

\end{thebibliography}

\end{document}